# Proximity induced room-temperature ferromagnetism in graphene probed with spin currents

J.C. Leutenantsmeyer*, A.A. Kaverzin*, M. Wojtaszek and B.J. van Wees

Physics of Nanodevices, Zernike Institute for Advanced Materials, University of Groningen, 9747 AG Groningen, The Netherlands

* These authors contributed equally to this work.

**The introduction and control of ferromagnetism in graphene opens up a range of new directions for fundamental and applied studies. Several approaches have been pursued so far, such as introduction of defects, functionalization with adatoms, and shaping of graphene into nanoribbons with well-defined zigzag edges.**[1–6] **A more robust and less invasive method utilizes the introduction of an exchange interaction by a ferromagnetic insulator (FMI) in proximity with graphene.**[7–14]

**Here we present a direct measurement of the exchange interaction in room temperature ferromagnetic graphene. We study the spin transport in exfoliated graphene on a yttrium-iron-garnet (YIG) substrate where the observed spin precession clearly indicates the presence and strength of an exchange field that is an unambiguous evidence of induced ferromagnetism. We describe the results with a modified Bloch diffusion equation and extract an average exchange field of the order of 0.2 T. Further, we demonstrate that a proximity induced 2D ferromagnet can efficiently modulate a spin current by controlling the direction of the exchange field. These results can create a building block for magnetic-gate tuneable spin transport in one-atom-thick spintronic devices.**[8,9]

The magnetic proximity effect describes the introduction of ferromagnetic order into an intrinsically nonmagnetic material by an adjacent ferromagnet. Being atomically thin, graphene presents an ideal platform for studying such interaction, in particular when combined with a ferromagnetic insulator. Theory predicts that for the idealised case of (super)lattice matching an exchange interaction of tens of meV can be obtained.[14,15] Up to date it has been studied experimentally in a number of FMI/graphene systems using materials with low Curie temperature such as EuO ($T_c = 69$ K) and EuS ($T_c = 16.5$ K).[11,12] In comparison, YIG provides the advantage of a Curie temperature of 550 K, along with chemical stability in atmospheric conditions, the preservation of the charge transport properties in the graphene and the possibility to directly exfoliate or transfer graphene onto the surface for fabricating graphene-based spintronic devices.

As indication of a ferromagnetic exchange interaction in graphene-YIG heterostructures the observation of an anomalous Hall effect was reported.[7] More recently the presence of an exchange interaction in YIG/CVD graphene devices was invoked to explain magnetoresistance measurements and ferromagnetic resonance spin pumping.[13] So far, in all the reports the authors employ charge transport, where in addition to exchange interaction also spin-orbit interaction is needed for the understanding, both a priori unknown parameters.

In this work, we probe the induced exchange interaction in graphene in the most direct way using only the spin degree of freedom. The magnetic interaction from the YIG is expected to produce an exchange term in the Hamiltonian and to spin split the graphene energy bands (Fig. 1a). It can be described as an additional effective field that is determined by the direction and magnitude of the YIG magnetization. By studying its effect on spin transport and precession, and fitting the results with the modified Bloch diffusion equation we are able to describe our results qualitatively and quantitatively. We further demonstrate that the exchange field can be used for an efficient modulation of spin currents.

The device is shown in Fig. 1b and c. A single layer graphene flake of approximately 12 μm by 1.2 μm is first exfoliated on $SiO_2$ and transferred to YIG. Ferromagnetic contacts are defined via e-beam lithography followed by Ti deposition, in-situ oxidation to form $TiO_2$, Co deposition and liftoff. A non-local spin valve characterization[16] is shown in Fig. 2. A charge current is sent from the injector to the reference electrode. As a result a pure spin current diffuses through the channel and is detected as a voltage difference between the detecting and another reference electrode. The spin-transport measurements are obtained using contacts 1 and 2 as injector and detector with contact spacing $d = 1.2$ μm. The magnetization direction of the injector (detector) can be controlled by sweeping the applied magnetic field along the easy axis of the electrodes. Fig. 2b shows the change of the non-local resistance ($R_{NL}$) when the electrode configuration is switched between parallel and antiparallel alignment. The change in $R_{NL}$ is a pure spin signal that increases from 90 mΩ at room temperature to 650 mΩ at 75 K. To determine the spin relaxation length λ, we fit the dependence of the spin signal on $d$ and extract $\lambda = (490 \pm 40)$ nm.[15] These values are comparable to our other graphene devices on YIG or $SiO_2$[15,17], which confirms that the basic spin transport properties of graphene are conserved after transfer to YIG.

To investigate the presence of the exchange interaction, we study the Hanle spin precession (Fig. 3a). A perpendicular magnetic field causes injected spins to precess while diffusing along the channel, changing the average polarization and direction of the spins at the detector. The total effective field ($\boldsymbol{B}_{tot}$) that is sensed by the spins consists of the applied field ($\boldsymbol{B}_{app}$) and exchange field ($\boldsymbol{B}_{exch}$), $\boldsymbol{B}_{tot} = \boldsymbol{B}_{app} + \boldsymbol{B}_{exch}$. $\boldsymbol{B}_{app}$ is swept perpendicular to the sample plane and causes the Zeeman splitting of the graphene spins, $\Delta E_{Zeeman} = g\mu_B|\boldsymbol{B}_{app}|$. The exchange field is determined by the YIG magnetization direction $\boldsymbol{M}$ and the interface properties and is defined as $\Delta E_{exch} = g\mu_B|\boldsymbol{B}_{exch}|$. Here $g$ is the gyromagnetic factor (~2 for graphene) and $\mu_B$ the Bohr magnetron.

Typical Hanle curves for graphene devices on $SiO_2$[18] or hBN[19] substrates are smooth in the full measured range, whereas we clearly observe a sharp transition at $\boldsymbol{B}_{app} \sim 180$ mT in our sample. The kink is seen for both parallel and antiparallel injector/detector magnetizations at all measured temperatures although it becomes more pronounced at 75 K. The appearance of such transition requires an additional spin precession caused by the exchange field $\boldsymbol{B}_{exch}$ that is constant in magnitude and collinear with $\boldsymbol{M}$. When no external field is applied, $\boldsymbol{M}$ together with $\boldsymbol{B}_{exch}$ lies within the sample plane and is gradually pulled out of the plane with increasing $\boldsymbol{B}_{app}$. The transition point coincides with the saturation field of the YIG magnetization ($B_s \sim 180$ mT) above which $M$ and $B_{exch}$ are aligned fully perpendicular to the sample plane. The change of the transition field with temperature is consistent with the change of the magnetization saturation field of the YIG films.[15] Thus, we conclude the existence of an exchange field with a magnitude comparable to the applied field at the transition point ($\sim 180$ mT).

To further confirm the presence and magnitude of the exchange interaction, we utilize the low in-plane coercivity of YIG. By applying and rotating a small magnetic field of 20 mT in the sample plane we can drag along the magnetization direction of the YIG while maintaining the parallel/antiparallel alignment of the injector/detector electrodes (Fig. 4a). Thus, we can rotate the direction of the exchange field while keeping the direction of the injected/detected spins unaffected. When $B_{exch}$ is collinear with the injected spin polarization ($\beta = \pm 90°$) it has no influence on the spin transport, whereas at $\beta = 0°$ diffusing spins experience the maximum precession and dephasing. In Fig. 4b the dependence of the spin signal on β is shown for both parallel and antiparallel magnetization alignment. For $d = 0.9$ μm (contacts 2 and 3, Fig. 1c) the observed modulation is around 50 %, which is substantial and cannot be explained by the effect of $B_{app} (= 20$ mT) alone. With increasing distance between both electrodes the modulation reaches $\sim 100\%$ at $d = 4.2$ μm (Figs. 4b-d). In this case all spin components perpendicular to $B_{exch}$ are dephased and averaged to zero and the spin signal has a dependence close to $\Delta R_{NL}(\beta) = \pm R_{NL}^0 \cos^2 \beta$. These measurements confirm that the magnitude of the exchange field corresponds to approximately 0.2 T.

To model the spin transport in graphene in the presence of an exchange interaction, we add the exchange field to the one dimensional Bloch equation:

$$0 = D_s \nabla \mu_s - \frac{\mu_s}{\tau_s} + \frac{g\mu_B}{\hbar}(B_{app} + B_{exch}) \times \mu_s,$$

where $D_s$ denotes the spin diffusion coefficient, $\mu_s$ the three-component spin chemical potential, $\tau_s$ the spin relaxation time and $\hbar$ the reduced Planck constant. We obtain an analytical expression for the spin accumulation at the detector depending on $B_{app}$.[15] Below the YIG saturation field $B_s$ both $M$ and $B_{exch}$ are determined by the standard easy-plane magnetic anisotropy model, whereas above the saturation field $\boldsymbol{B}_{exch}$ is fixed and aligned with $\boldsymbol{B}_{app}$.[15] We use our model to qualitatively reproduce the observed behaviour and also make a quantitative estimation of the exchange field. The fitting of the Hanle curve is shown in Figs. 5a-b, where we find the best agreement for $\lambda = 1.8\ \mu m$ ($\lambda = \sqrt{D_s \tau_s}$) and $|\boldsymbol{B}_{exch}| = 0.2\ T$. A possible explanation for the difference between $\lambda$ extracted from the distance dependent spin signal ($\lambda = 490\ nm$) and from the modeling ($\lambda = 1.8\ \mu m$) is a spatial inhomogeneity of the exchange field. $|\boldsymbol{B}_{exch}|$ is expected to depend crucially on the overlap between the graphene π-orbitals with the iron d-orbitals of the FMI. One can readily assume that the regions where the electrodes are on top of the graphene experience a different exchange interaction then regions where the graphene lies freely on the surface.

Our analysis can be further extended to the spin transport modulation dependencies shown in Fig. 4b-d. Fig. 5b shows the modulation of the spin signal depending on $d$. These results can also be fit with $\lambda = 1.8\ \mu m$ and $|\boldsymbol{B}_{exch}| = 0.2\ T$. When extrapolating the data to $d = 0$, we find ~35% modulation which can be obtained analytically from our model:

$$\left.\frac{R_{NL}^{max} - R_{NL}^{min}}{R_{NL}^{max}}\right|_{d=0} = 1 - \frac{1}{\sqrt{2}} \frac{\sqrt{1 + \sqrt{1 + (\omega \tau_s)^2}}}{\sqrt{1 + (\omega \tau_s)^2}}$$

where $\omega = \frac{g}{\hbar} \mu_B |\boldsymbol{B}_{exch}|$. Using $|\boldsymbol{B}_{exch}| = 0.2\ T$, we obtain $\tau_s \sim 40\ ps$. Assuming that the spin and charge diffusion ($D_c$) coefficients coincide we deduce $\lambda = \sqrt{D_c \tau_s} \sim 500\ nm$, which resembles the $\lambda$ extracted from distance dependent spin signal, again suggesting an inhomogeneous exchange field.[15]

In summary, we have demonstrated the detection of ferromagnetic exchange interaction in graphene by spin transport at room temperature, 75 K and 4.7 K.[15] The exchange field strength is quantified to be 0.2 T. Given the theoretical results on idealised systems, substantial enhancement should be possible by appropriate interface optimisation.[14] We proposed spin-transport measurements as the most direct way to study the exchange field in graphene. Furthermore, we showed that a spin current can be efficiently modulated at room temperature by controlling the exchange field, which opens up new directions to control spins in graphene based spintronic devices.

## Methods

Our graphene flakes are exfoliated from HOPG graphite crystals (HQ Graphene) on silicon oxide substrates. Single layer graphene flakes are selected by optical contrast and transferred to target substrates with a custom-built transfer stage using a polycarbonate based pickup technique. The commercially available (111) single crystal YIG films (Matesy GmbH) are grown by liquid phase epitaxy with 210 nm YIG thickness on GGG substrates. The films are cleaned with acetone, isopropanol and 180 s in 200 W oxygen plasma to remove organic residues. To minimize water contamination at the interface between graphene and YIG, the substrates are kept for 15 min in a furnace at 500°C until the graphene transfer at 140°C.

After transfer, the polycarbonate is dissolved in chloroform and the graphene is cleaned for one hour in a furnace at 350°C in an Ar/$H_2$ atmosphere. The flake is connected with electrodes made of titanium oxide tunnel barriers (0.8 nm), ferromagnetic cobalt electrodes (45 nm) and an aluminium capping layer (5 nm) using e-beam lithography. The samples are characterized in a cryostat with standard AC-lock-in measurement techniques at room as well as low temperatures. We apply typical AC currents between 1 and 20 µA with frequencies between 1 and 13 Hz. At 75K the electrodes show a contact resistance of the order of 1 – 3 kΩ and a spin signal between 7 Ω at 500 nm contact spacing and 10 mΩ at 3.9 µm spacing. Measurements of Shubnikov-de Haas oscillations at T = 2 K, reveal a carrier density of the order of $3*10^{12}$ $cm^{-2}$ and a mobility of 720 $cm^2$/Vs. In different graphene/YIG samples we observe holes as charge carriers resulting from doping during the transfer process. Further details are discussed in the supplementary information.

## Acknowledgements

We acknowledge A. Aqeel, P. Zomer, M. de Roosz and J.G. Holstein for technical assistance and J. Ingla-Aynes and A.M. Kamerbeek for fruitful discussions. The research has received funding from the European Union's 7[th] Framework Program within the Marie Curie initial training network 'Spinograph' (grant 607904), the 'Graphene Flagship' (grant 604391) and the Dutch 'Foundation for Fundamental Research on Matter' (FOM).

## Author contributions

BJvW, JCL, AAK and MW conceived the experiments. JCL and AAK designed and carried out the experiments. JCL, AAK and BJvW analysed and discussed the data and wrote the manuscript.

## Competing financial interests

The authors declare no competing financial interests.

# List of figures

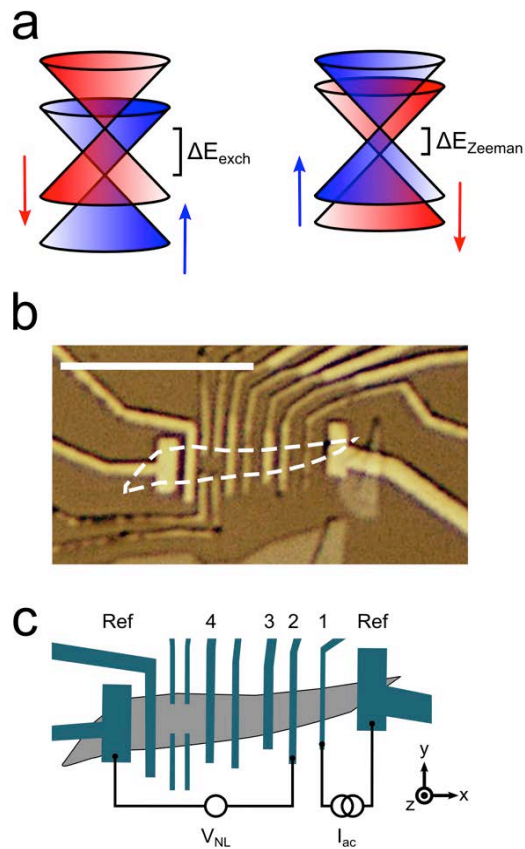

**Figure 1: Spin transport in graphene in the presence of a proximity induced exchange field.**

(a) The exchange field creates a splitting of the Dirac cone for each spin species similar to the magnetic field induced Zeeman splitting. Note that both the sign and magnitude of the exchange and Zeeman splitting can be different. (b) Optical micrograph of the graphene/YIG heterostructures indicating the single layer graphene flake and the deposited $TiO_2$/Co contacts. To control the coercive field of the electrodes for magnetization switching, we define different contact widths between 200 and 400 nm. The scale bar represents 10 µm. (c) Schematic sketch of the sample showing the characterized injector/detector contacts (1, 2, 3, 4) and reference contacts (Ref). The circuit for the measurements shown in Fig. 2 and 3 is indicated.

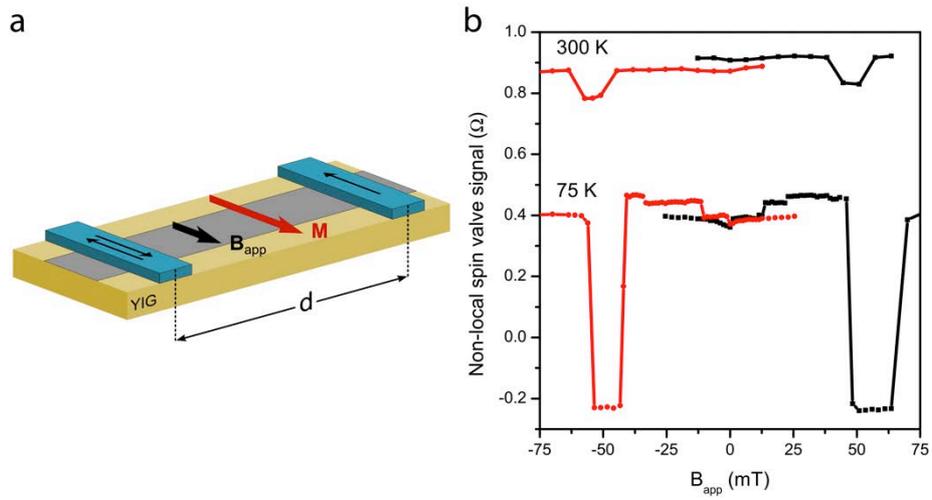

**Figure 2: Non-local spin transport in graphene spin valves on YIG.** (a) Schematic measurement to characterize the non-local spin transport by switching the electrode magnetization with an external magnetic field ($\boldsymbol{B}_{app}$). $\boldsymbol{M}$ denotes the magnetization of the YIG film and is parallel to $\boldsymbol{B}_{app}$ for these measurements. (b) The experimental data of the spin valve switching at 300 K and 75 K is measured with $d = 1.2$ µm. The black squares (negative to positive) and the red circles (positive to negative) correspond to different sweep directions of $\boldsymbol{B}_{app}$. The other switches can be attributed to smaller contributions of the outer electrodes to the spin signal.

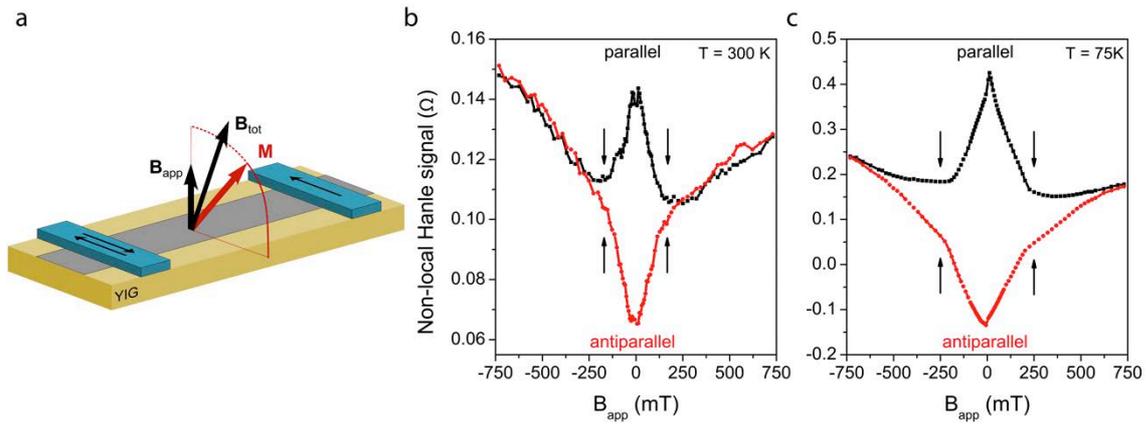

**Figure 3: Hanle spin precession modified by the proximity exchange field.** (a) The schematic measurement setup. $B_{app}$ is applied perpendicular to plane to cause Hanle spin precession in the graphene transport channel. The spins precess around the total effective field $B_{tot}$. (b) Data at room temperature and 75 K (c). The top curve (black squares) indicates the signal measured with parallel injecting and detecting electrodes, the bottom curve (red circles) represent the antiparallel alignment. The black arrows indicate the perpendicular saturation field of the underlying YIG film.[15]

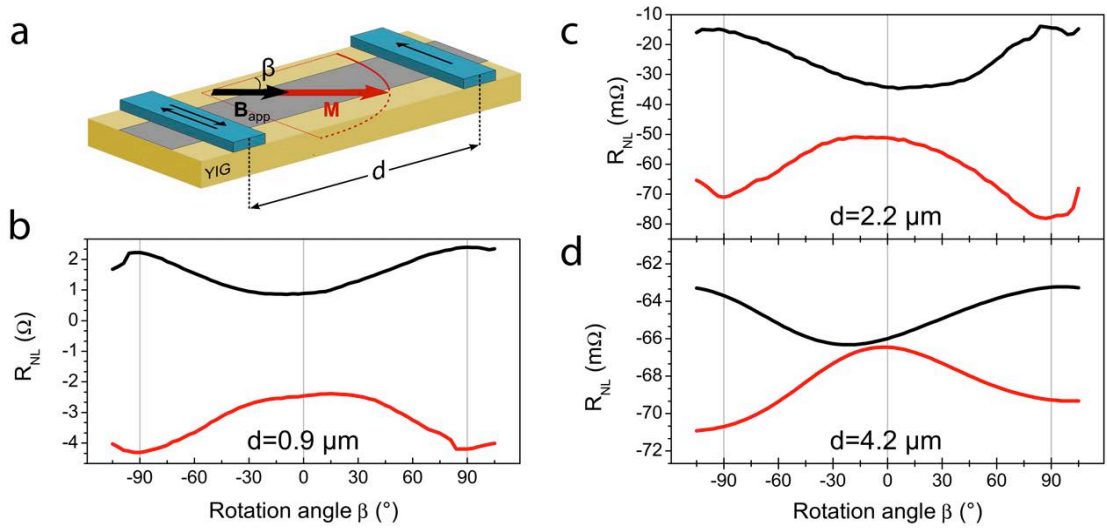

**Figure 4: Modulated spin transport with an in-plane exchange field.** (a) Schematic of the measurement, the YIG magnetization is rotated by $|B_{app}| = 20$ mT by an angle β with parallel (black) and antiparallel (red) electrode alignment. The rotation causes a modulation of the spin signal, which is shown for three different distances (b-d). For the farthest distance a smoothed curve is plotted. The relative modulation increases from 50% at $d = 0.9$ µm up to ~100% at $d = 4.2$ µm injector to detector spacing.

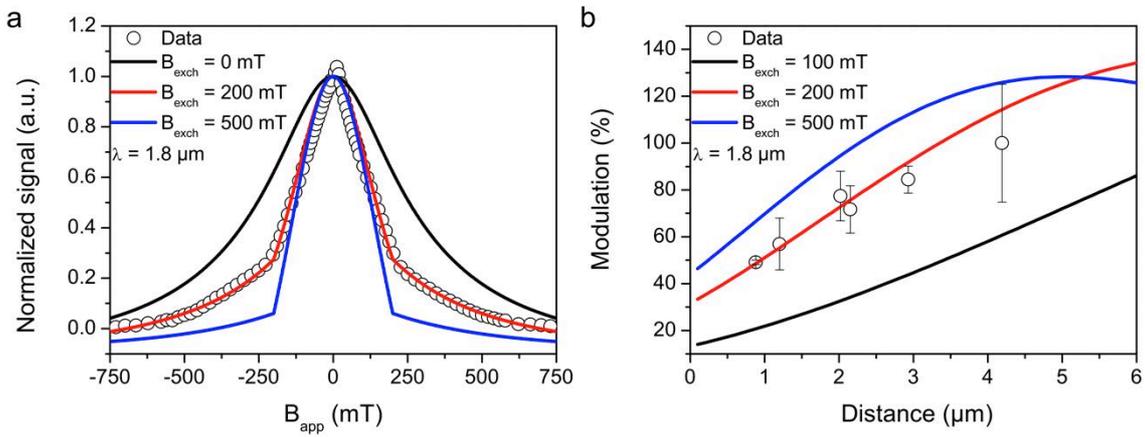

**Figure 5: Modelling of the spin transport in the presence of an exchange field.** (a) The standard spin transport model is modified to take the proximity induced exchange field into account. The resulting Hanle spin precession curve in a perpendicular $\boldsymbol{B}_{app}$ is shown for three different exchange fields and compared with the measured data. (b) The modulation of the spin signal originating from the in-plane exchange field is shown for three different exchange fields. In both independent measurements we extract $|\boldsymbol{B}_{exch}| \sim 0.2$ T from the data.

# Supplementary information for 'Proximity induced room-temperature ferromagnetism in graphene probed with spin currents'

## I. Characterization of the YIG/GGG films

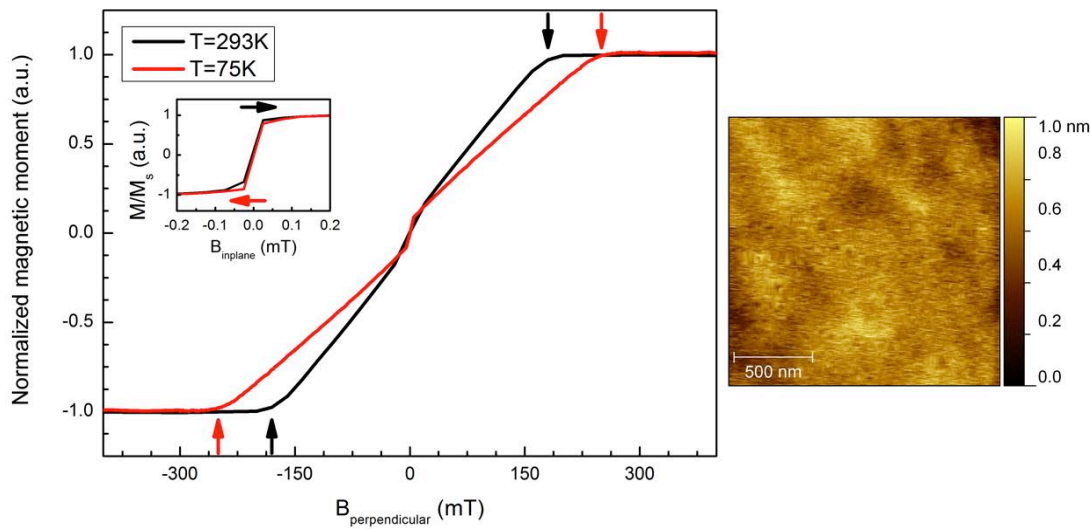

**Figure 1:** Magnetisation response of the 210 nm (111) YIG film on applied field and the surface topography. left: the YIG film magnetisation in an applied perpendicular magnetic field at 293 K (black) and 70 K (red). In accordance with an easy-plane magnetisation

anisotropy model it can be described as $M_Z = |M| \cdot B_{app}/B_S$ below $B_S$ (see main text for the notations). The magnetisation $M_Z$ saturates at around $B_S$ ~180 mT at room temperature and at 250 mT at 75 K. The inset shows the in-plane magnetization measured at 293 K showing an in-plane coercive field of less than 0.1 mT. Right: The AFM image shows a YIG substrate cleaned before the graphene transfer as described in the methods section. The topography has an RMS surface roughness of the order of 0.1 nm over a 1.75 μm by 1.75 μm square. Smaller scale corrugations (~20 nm in lateral dimension) are due to the line scanning of the image. The lateral resolution of the AFM scan is ~10 nm.

## II. Charge and spin transport properties of the graphene flake

The thickness of the graphene flake is determined after exfoliation on 300 nm $SiO_2$ on a Si substrate by optical contrast (on average around 6% per single layer in our system). The discussed flake has a contrast of 5.5% from which we conclude single layer thickness. To estimate the carrier concentration we use the Shubnikov-de Haas oscillations of the longitudinal resistance at 2.2 K (Fig. 2). The reciprocal magnetic field values of the minima are plotted as a function of the peak index and shown in the inset of Fig. 2. We use only the three highest field minima to assure an accurate estimation of the carrier density. The $1/B$ slope of 0.032 $T^{-1}$ corresponds to a carrier density of $n = 3 \cdot 10^{12}$ $cm^{-2}$. This value is in good agreement with our other similarly fabricated samples that show holes as carriers with densities of $n \sim 10^{12}$ to $10^{13}$ $cm^{-2}$.

Using the resistance of region 1, we deduced a carrier mobility of 720 $cm^2/Vs$, also a typical value for our other graphene/YIG devices. With the obtained carrier density, we calculate a charge diffusion coefficient of $D_c = 66$ $cm^2/s$.

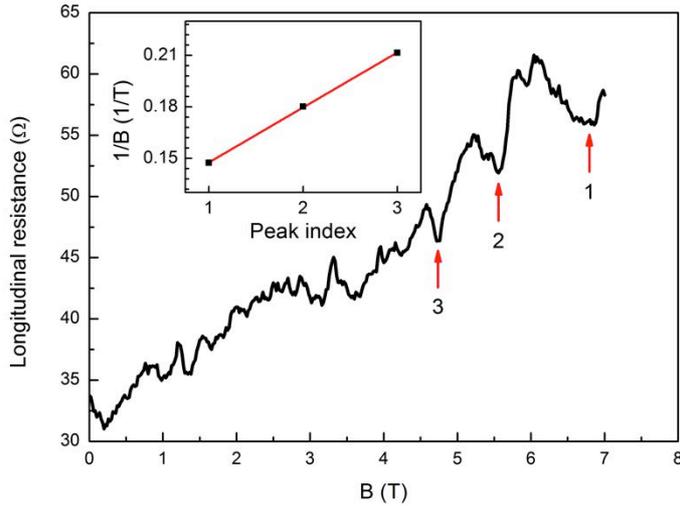

**Figure 2:** The longitudinal resistance of region 1 in high fields at T = 2.2 K. We observe three minima of the Shubnikov-de Haas oscillations. The inset contains the reciprocal field position of the three highest minima as a function of the peak index.

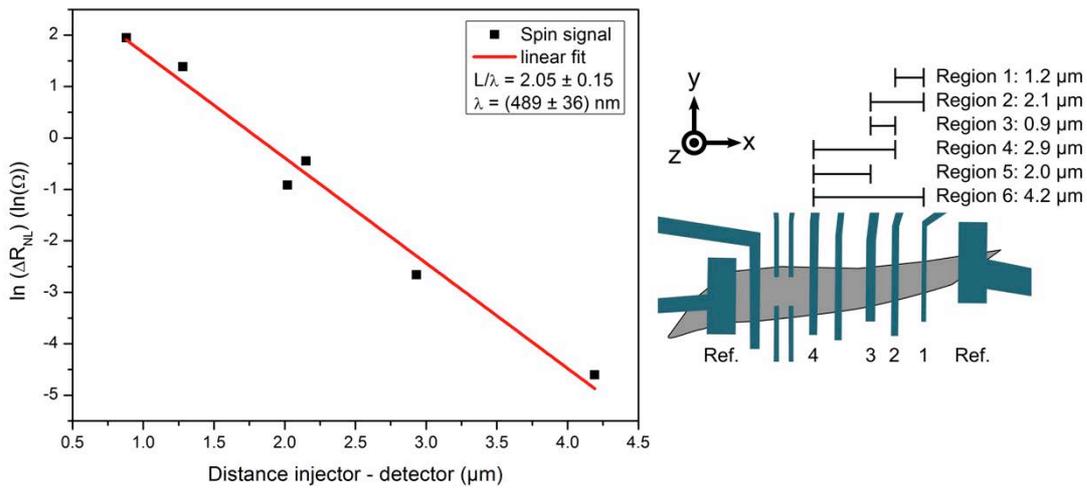

**Figure 3:** Left: The distance d dependent spin signal is shown and used to calculate the spin relaxation length λ. Right: Schematic image of the graphene flake and patterned contacts. The contacts and pairs of injector-detector with different d used for measurements are labelled and axes defined.

From the distance dependent spin signal (Fig. 3, left) we obtain a spin relaxation length λ = (490 ± 40) nm, which is in agreement with our previous samples of graphene/YIG heterostructures (λ ~ 700 nm).

# III. Characterization of the graphene flake and cobalt electrodes

| Region | $R_i$ (kΩ) | $R_d$ (kΩ) | $R_c^{eff}$ (kΩ) | l×w (μm$^2$) | R (kΩ) | $R^{sq}$ (kΩ) | r (μm) | r/λ |
|---|---|---|---|---|---|---|---|---|
| 1 | 1.9 | 0.89 | 1.212 | 1.2 × 1.1 | 6.1 | 5.6 | 0.238 | 0.49 |
| 2 | 1.9 | 0.92 | 1.240 | 2.1 × 1.3 | 12.8 | 7.9 | 0.203 | 0.42 |
| 3 | 0.98 | 0.92 | 0.905 | 0.9 × 1.4 | 6.6 | 10.3 | 0.123 | 0.25 |
| 4 | 0.98 | 0.65 | 0.751 | 2.9 × 1.4 | 15.2 | 7.4 | 0.143 | 0.29 |
| 5 | 0.92 | 0.65 | 0.762 | 2.0 × 1.6 | 8.5 | 6.5 | 0.188 | 0.38 |
| 6 | 1.90 | 0.65 | 0.969 | 4.2 × 1.4 | 21 | 7.0 | 0.194 | 0.40 |

**Table 1**. Measured and derived parameters of the graphene flake and contacts. The table gives an overview of injector and detector resistances $R_i$ and $R_d$ measured in a 3-terminal geometry, effective contact resistance $R^{eff} = 2/(1/R_i + 1/R_d)$, length and width of the flake region l and w, resistance R of the flake region measured in 4-probe, calculated graphene square resistance $R^{sq} = R \cdot w/l$ and conductivity mismatch parameter $r = R^{eff} / R^{sq} \cdot w$ with ratio r/λ.

To analyse the influence of the contacts on the spin relaxation length λ, we calculate the conductivity mismatch parameter r for the different regions. [1,2] In the case when the contacts have the biggest contribution in spin relaxation, when r/λ = 0.25, the intrinsic spin relaxation length can be higher than the extracted value by ~ 20%.

# IV. Discussion of the full data sets. Effect of the cobalt stray field on YIG magnetisation direction

We measured the spin precession at 4.7 K, 75 K and 300 K and extracted the spin signal by subtraction of the antiparallel from the parallel Hanle curve. The amplitude of the spin signal is observed to increase with decreasing temperature, which is also consistently seen in the spin valve measurements. The characteristic kink at the saturation field of the YIG magnetization as well as the relatively linear shape of the Hanle below saturation field and the remaining spin signal up to about 600 mT are present at all temperatures.

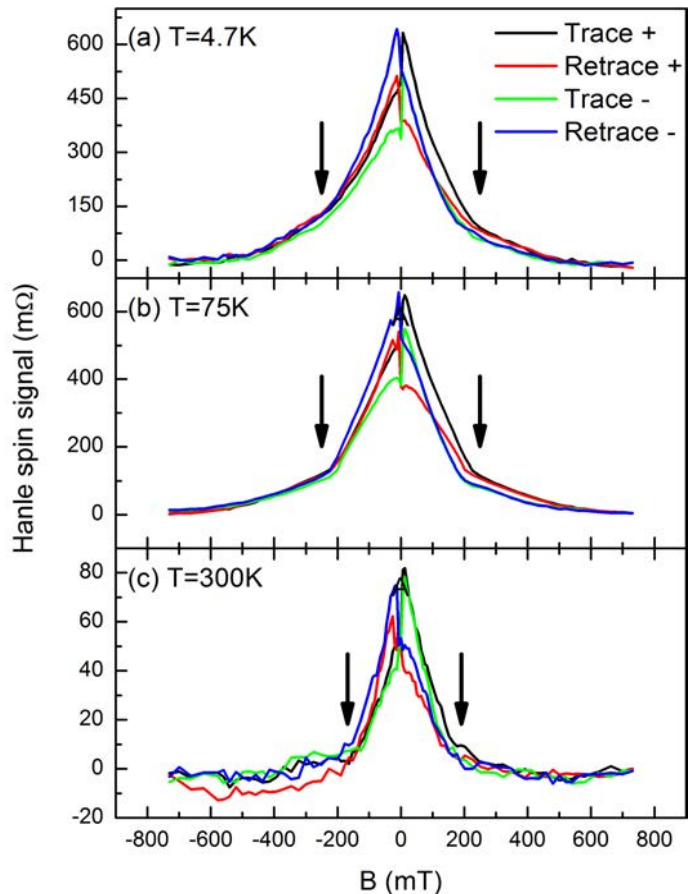

**Figure 4:** Magnetic field dependence of the spin signal obtained from Hanle precession data sets at different temperatures. Arrows indicate the YIG saturation fields.

We find the kink to shift from 180 mT at room temperature to higher fields with decreasing temperature, which is in agreement with the SQUID measurements of the YIG films (see section I). To extract the spin dependent signal, we subtract the Hanle curves measured in parallel and antiparallel configurations as plotted in Fig. 4. The + designates the difference between parallel up-up and antiparallel up-down alignment of the injector/detector electrodes. The - denotes the difference between parallel down-down and antiparallel down-up alignment. Alignment up indicates that the contact magnetization is aligned along the y-axis (Fig. 3), in the positive magnetic field direction. Alignment down indicates the opposite, negative field, direction along the y-axis. All four curves (including trace and retrace curves for alignments + and -) show the transition point where the magnetization of the YIG film saturates. The switches of the spin signal close to zero applied field are due to the fact that the YIG magnetization around $B_{app} = 0$ cannot be well controlled, leading to an abrupt change of the in-plane magnetization direction and resulting in a change of the exchange field acting on the spins.

This can be explained with the stray field arising from the contact magnetization. The coercive field of the YIG magnetization in the film plane (easy-plane) is smaller than 0.1 mT (see section I). Thus, even a rather small stray field can locally influence M. From the given geometry of the contacts we conclude that the strongest stray field is expected at the ends of the electrodes, which is typically more than 1 µm away from the graphene channel. Therefore, the direct contribution from the stray field of the contacts to the field acting on graphene is negligible, which we confirmed with finite element modelling in COMSOL multiphysics. However, at small applied fields the YIG magnetization can be influenced by the contact alignment which, thus, determines the exact switching behaviour of M. Moreover, it is seen that the switching of Trace + and Retrace - or Trace - and Retrace + is symmetric with respect to the zero field, which can be understood by taking the magnetization direction of the contacts into account.

## V. Possible effect of stray fields on graphene

As discussed in section IV the stray fields arising from the contact magnetization are expected to be negligible close to the graphene flake and cannot directly affect the spin transport in the channel. Another possible source of stray fields is the YIG film in direct vicinity to the graphene. Assuming a perfect flatness of the 210 nm YIG and a typical film size of 5 × 5 mm, no stray fields are expected. However, it was shown in by Dash et al. [3] that the finite roughness of the surface of the magnetic material can lead to non homogeneous in-plane and out-of-plane stray fields. Under assumption of perfectly flat graphene, out-of-plane stray fields average out spatially as the total magnetic flux through the graphene surface has to be zero. With finite roughness of graphene both in-plane and out-of-plane fields can have a non-zero average value.

In our case the roughness of the YIG film is ~ 0.1 nm (section I), therefore, we expect a negligible magnitude of a YIG stray field. Moreover, based on our analysis we can exclude the effect of stray fields because of the following reasons. Firstly, to explain our results with stray fields only, a magnitude of the order of 0.2 T, comparable to the YIG saturation field, would be required. Such a large average stray field cannot originate from the measured YIG roughness. Secondly, fitting of both out-of-plane Hanle measurements (Fig. 3 main text) and in-plane rotation measurement (Fig. 4, main text) leads to a similar magnitude of the exchange field which therefore cannot be explained by stray fields, as they are expected to be very different for in-plane or out-of-plane magnetization configurations. Thus, we conclude that possible stray fields from YIG do not affect our measurements.

## VI. Influence of a magnon transport channel

As shown by Cornelissen et al. [4] and Goennenwein et al. [5] a spin accumulation induced in a material in proximity with YIG can excite a magnon current in the FMI, leading to a parallel spin transport channel from the injector to the detector. We can exclude the existence of an additional spin transport channel in our graphene/YIG system, since the magnon transport process is suppressed at low temperatures [5], while we find an increase of the spin signal, confirming that the signal is not carried by magnons.

## VII. Comparison to a reference sample

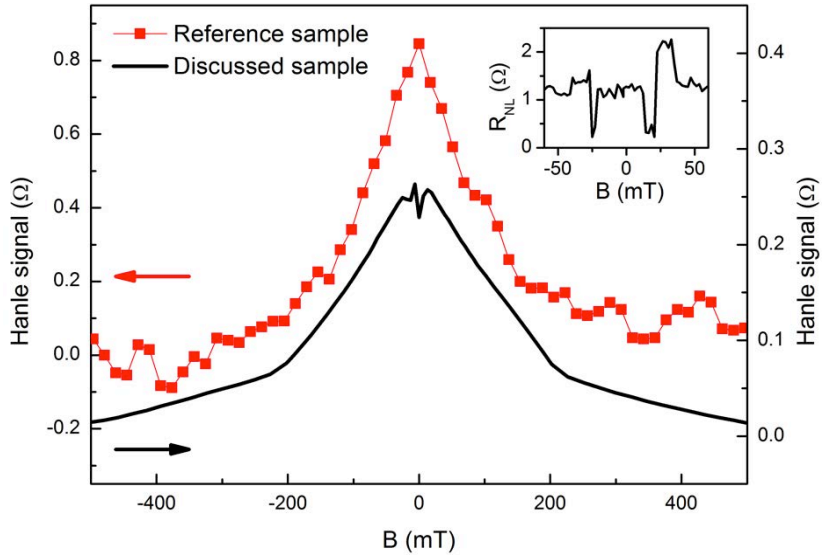

**Figure 5:** Comparison to a reference sample. The spin precession data of the additional reference sample (red squares, 1 µm contact spacing, 70K) is plotted together with the same measurement for main text sample (black line, 1.2 µm contact spacing, 75 K). Despite similar fabrication steps, we find a higher contact impedance and an increased noise level in the reference sample. However, we are able to observe a comparable magnitude of the spin signal as well as the characteristic features like the relatively linear shape at lower fields and the kink at the perpendicular saturation field of the substrate. The inset contains a non-local spin valve measurement of the reference sample.

# VIII. Full set of spin transport modulation data

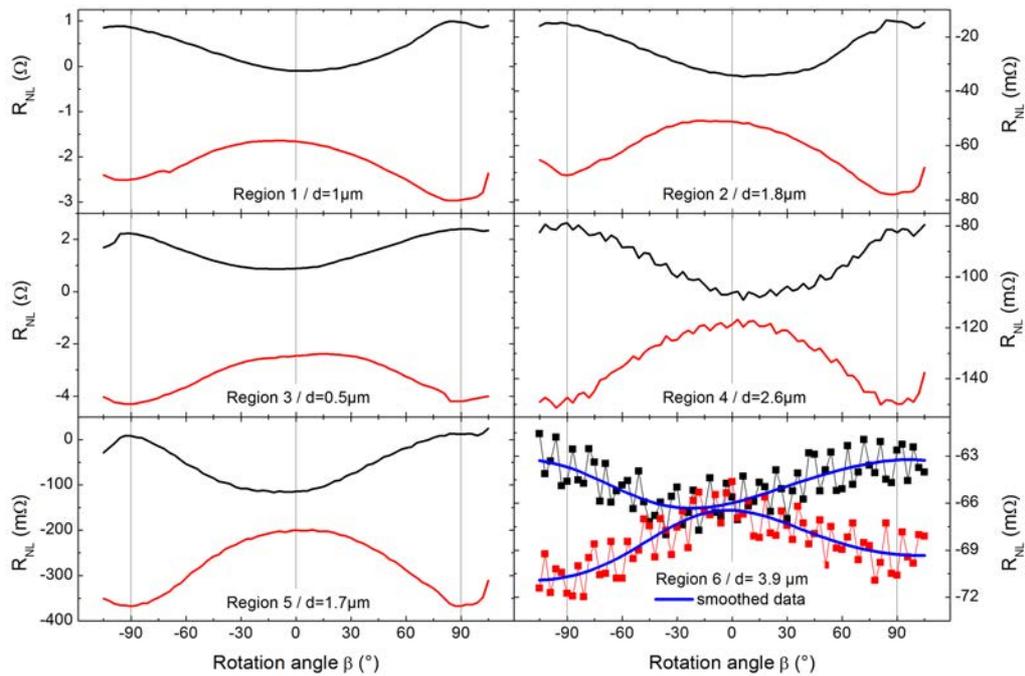

**Figure 6:** Full data set of the spin transport modulation by in-plane rotation of the magnetization direction at 75 K. The extracted relative modulation is discussed in the main text. For the farthest distance the raw data and the smoothed curves are shown.

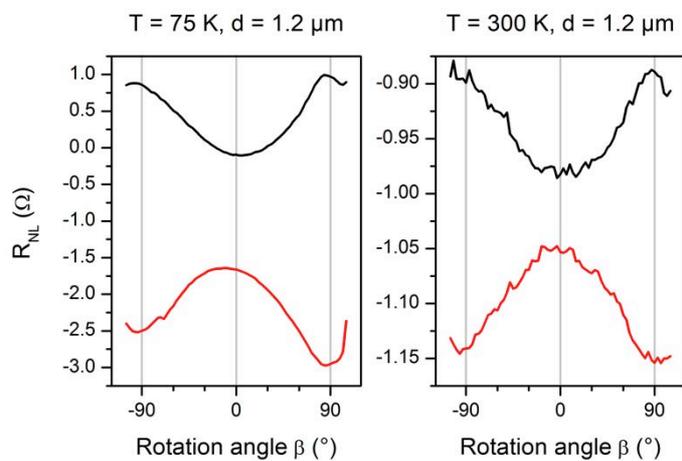

**Figure 7:** Spin modulation in region 1 at room temperature and 75 K. The spin signal in region 1 (1.2 µm) at 75 K is modulated by 57%. The modulation increases at room temperature to 77% (100 mΩ).

## IX. Modelling

To model the observed data we solve the one dimensional Bloch diffusion equation for a total effective field acting on the spins:

$$0 = D_s \nabla \boldsymbol{\mu}_s - \frac{\boldsymbol{\mu}_s}{\tau_s} + \frac{g\mu_B}{\hbar}(\boldsymbol{B}_{app} + \boldsymbol{B}_{exch}) \times \boldsymbol{\mu}_s,$$

where $\mu_s(x) = (\mu_x(x), \mu_y(x), \mu_z(x))$ and the magnetic field $\boldsymbol{B}_{tot}$ is the vector sum of the external applied magnetic field and exchange field ($\boldsymbol{B}_{tot} = \boldsymbol{B}_{app} + \boldsymbol{B}_{exch}$). The equation can be solved with the boundary condition for the spin accumulation μs(x) = (0,0,0) at x = ±∞ and the assumption that the spins are injected only in y direction, $\frac{\partial}{\partial z}\mu_s(x) \sim (0, \mu_y^0, 0)$ at x = 0.

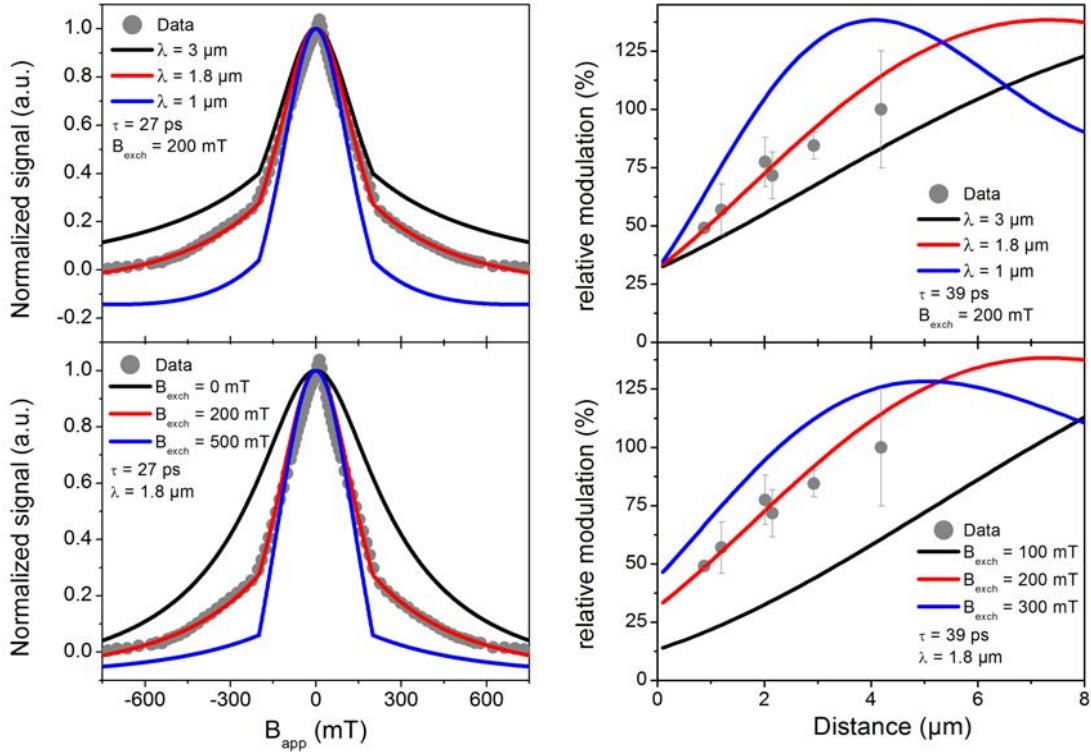

**Figure 8:** Influence of the spin relaxation length and exchange field on the Hanle spin precession and spin signal modulation. The derived model is used to fit experimental results. The circles represent the experimental results and solid lines the model curves for different parameters. Left: fitting of the Hanle dependence (Fig. 3 main text). Right: fitting of the relative modulation dependence derived from Fig. 4 (main text).

The analytical solution for the y-component of the spin accumulation reads:

$$\mu_y(x) = A \left[ \frac{B_y^2}{B^2} e^{-\frac{x}{\lambda}} + \frac{B^2(B_x - B_z)^2 + (B_x^2 - B_xB_y - B_zB_y + B_z^2)^2}{2B^2(\alpha_1^2 + \alpha_2^2)(B^2 - B_xB_y - B_yB_z - B_zB_x)} \right.$$

$$\left. \left( \alpha_1 \cos\left(\alpha_2 \frac{x}{\lambda}\right) - \alpha_2 \sin\left(\alpha_2 \frac{x}{\lambda}\right) \right) e^{-\alpha_1 \frac{x}{\lambda}} \right]$$

where $A$ is a scaling parameter, $\alpha_{1(2)} = \frac{1}{\sqrt{2}}\sqrt{\pm 1 + \sqrt{1 + (\omega\tau_s)^2}}$, $\omega = \frac{g}{\hbar}\mu_B B = \frac{g}{\hbar}\mu_B |\mathbf{B}_{tot}|$ and $\tau_s = \lambda^2/D_s$.

For relevant cases when either $B_x = 0$ or $B_y = 0$ the expression for $\mu_y(x)$ can be simplified:

$$\frac{B^2(B_x - B_z)^2 + (B_x^2 - B_xB_y - B_zB_y + B_z^2)^2}{2B^2(B^2 - B_xB_y - B_yB_z - B_zB_x)} = \begin{cases} \left(\frac{B_z}{B}\right)^2, & \text{when } B_x = 0; \\ 1, & \text{when } B_y = 0. \end{cases}$$

The obtained expression μ$_y$(x) is used to fit three types measurements. First, we fit the Hanle precession data when the external field is applied perpendicular to the sample plane (Figs. 3c and 5a from the main text). A, λ, τ$_s$ and $|\boldsymbol{B}_{exch}|$ are used as parameters. The best fit is obtained with $|\boldsymbol{B}_{exch}| \sim 0.2$ T , $\tau_s \sim 27$ ps and λ ~ 1.8 μm, Fig. 8a-b. From the modelling we conclude that when M is aligned with B$_{app}$, B$_{app}$ and B$_{exch}$ have the same sign.

Secondly, we fit the relative modulation of the spin signal as a function of the distance between electrodes when the magnetization of the YIG is rotated in the sample plane (Fig. 4, main text) with A, λ and $\tau_s * |\boldsymbol{B}_{exch}|$. The best fit is obtained with λ ~ 1.8 μm and $\tau_s * |\boldsymbol{B}_{exch}| = 7.8$ ps $*$ T, $|\boldsymbol{B}_{exch}| \sim 0.2$ T and $\tau_s \sim 39$ ps, Fig. 8 c-d.

Lastly, we fit the in-plane Hanle precession when the direction of applied field is fixed along the x-axis, within the sample plane but perpendicular to the easy-axis of the contacts magnetization (Fig. 9). Due to smaller in-plane shape anisotropy compared to out-of-plane, we take the rotation of the contact magnetization in-plane into account. While the out-of-plane saturation field is around 1.2 T, the in-plane saturation field along the x-axis is between 100–200 mT, leading to a deviation of the direction of the injected spins relative to the y-axis. The contact magnetization and, therefore, the spin accumulation is calculated with an easy axis magnetic anisotropy model using the saturation field of the contact magnetization as an additional parameter. With the previously obtained values λ ~ 1.8 μm and $|B_{exch}|$ ~ 0.2 T we can qualitatively fit the measured dependencies with $\tau_s$ ~ 15 ps and a contact saturation field of 140 mT. It implies that above 140 mT both contact magnetisations and the YIG magnetisation are aligned with the external magnetic field, leading to a maximum spin signal. However, we find a further increase of the spin signal of unknown origin when applying magnetic fields up to 7 T (Fig. 9).

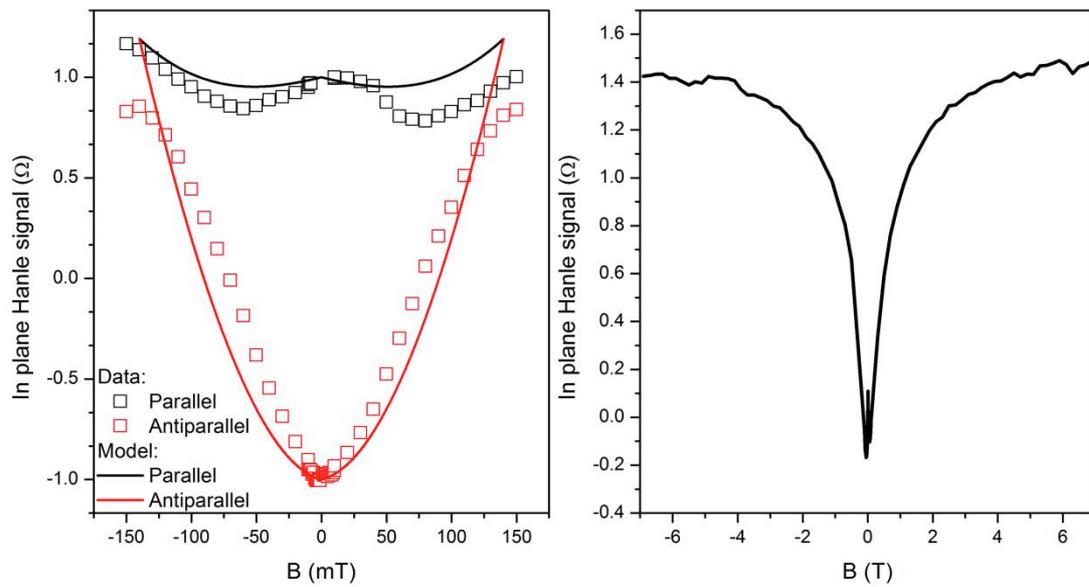

**Figure 9:** In-plane Hanle spin precession. Left: Hanle curves with the magnetic field applied in-plane along x-axis for parallel (black squares) and antiparallel (red squares) alignments fitted with the model (solid lines) taking into account in-plane rotation of the contact magnetization with applied field. Right: Measurement of the in-plane Hanle curve up to 7 T.